\documentclass{article}

 \usepackage[preprint]{neurips_2026}

\usepackage[utf8]{inputenc} 
\usepackage[T1]{fontenc}    
\usepackage{hyperref}       
\usepackage{url}            
\usepackage{booktabs}       
\usepackage{amsfonts}       
\usepackage{nicefrac}       
\usepackage{microtype}      
\usepackage{xcolor}         

\usepackage[space]{grffile}
\usepackage{amssymb} 
\usepackage{mathtools}
\usepackage{wrapfig}

\usepackage{float}              
\usepackage{algorithm}
\usepackage{algpseudocode} 
\usepackage{booktabs}
\usepackage{multirow}

\newcommand{\argmax}{\operatorname*{arg\,max}}

\newcommand{\GameState}{\mathbf{C}}
\newcommand{\Players}{N_P}

\title{A Benchmark for Multi-Party Negotiation Games from Real Negotiation Data}

\author{%
  Leo Benac\thanks{Equal contribution.} \\
  School of Engineering and Applied Sciences\\
  Harvard University\\
  \texttt{lbenac@g.harvard.edu} \\
  \And
  Jonas Raedler\footnotemark[1] \\
  School of Engineering and Applied Sciences\\
  Harvard University\\
  \texttt{jraedler@fas.harvard.edu} \\
  \And
  Zilin Ma \\
  School of Engineering and Applied Sciences\\
  Harvard University\\
  \texttt{zilinma@g.harvard.edu} \\
  \And
  Finale Doshi-Velez \\
  School of Engineering and Applied Sciences\\
  Harvard University\\
  \texttt{finale@seas.harvard.edu} \\
}

\begin{document}

\maketitle

\begin{abstract}
Many real-world multi-party negotiations unfold as sequences of binding, action-level commitments rather than a single final outcome, yet this regime remains under-studied in existing benchmarks. We introduce a benchmark and evaluation framework for this setting, combining a configurable negotiation game generator with document-grounded instances derived from a climate negotiation exercise. We also provide several baseline solvers. Exact evaluation on small games and comparative evaluation on larger instances show that no solver dominates across regimes; performance depends on the structural properties of the game. These results motivate the creation of novel negotiation methods that value partial commitments robustly across diverse strategic regimes. Code and data for the benchmark are available at: \url{https://anonymous.4open.science/r/negotiation_MARL-46B8/}.
\end{abstract}

\section{Introduction}

Computational models of negotiation have long been studied in game theory and multi-agent reinforcement learning \citep{rubinstein1982perfect, baarslag2010, abdelnabi2024cooperationcompetitionmaliciousnessllmstakeholders, mak2023coalitionalbargainingreinforcementlearning, meta_cicero}, with the promise of helping bounded decision-makers identify effective strategies in mixed-motive settings. Much of this work evaluates negotiation as bargaining over a specific outcome, typically under protocols where offers remain non-binding until acceptance \citep{lewis2017dealdealendtoendlearning, baarslag2010}. While this setting has yielded important insights, it leaves open how to evaluate negotiation in settings where intermediate commitments are binding and reshape future bargaining opportunities.

Indeed, many high-stakes, real-world multi-party negotiations cannot be reduced to a single final agreement. In our collaborations with professional frontline negotiators, we find that agreements are often built incrementally through sequences of bilateral and small-group deals in which parties make binding commitments over concrete actions. Each deal constrains future options and reshapes subsequent bargaining opportunities. Here, negotiation strategy includes not just the final agreement, but also the partner order, commitment sequencing, and how intermediate commitments interact across goals. However, because this type of negotiation remains under-evaluated, it is unclear whether existing methods capture these more granular strategic considerations (e.g., which parties to engage first, or which commitments are important to make early).

To address this evaluation gap, we introduce a benchmark and evaluation framework for sequential multi-party negotiation with binding commitments and terminal-only rewards, grounded in insights from real frontline negotiation practice. Beyond providing a new benchmark, our goal is to make explicit the assumptions involved in evaluating long-horizon negotiation under binding commitments, particularly how partial commitments are valued and how interaction is modeled. Specifically, our contributions are:

\textbf{Contribution 1: A benchmark for evaluating sequential negotiation under binding commitments.}
We introduce a benchmark for sequential multi-party negotiation games with binding commitments, grounded in real negotiation practice and data. The benchmark combines a configurable game generator with instances derived from climate negotiation position papers from a structured negotiation exercise. By varying structural properties such as incentive alignment, payoff structure, goal complexity, and goal non-linearity, it supports controlled evaluation across qualitatively different negotiation regimes.

\textbf{Contribution 2: A reference protocol and evaluation framework for long-horizon negotiation.}  
We provide a simple \emph{reference} negotiation protocol -- including turn-taking, partner selection, proposal construction, and acceptance -- that instantiates one natural way to interact with the benchmark. Using this fixed procedure, we enable controlled evaluation of negotiation models \emph{given} a negotiation protocol, reporting aggregate payoff improvements on large instances and exact deviations from optimal play on small ones. At the same time, the benchmark itself remains modular: the same game instances can be paired with alternative protocols, information structures, and objectives. Because these choices change what behavior is being evaluated, performance should be compared only within a fixed configuration and not across different protocol instantiations.  

\textbf{Contribution 3: A study of evaluative assumptions through value-function lenses}  
Among our baselines, we provide three simple value-function approximations, each encoding a different assumption about how the rest of the game will unfold. 
Across synthetic sweeps and document-grounded instances, we find that no approximation dominates across regimes and that different structural conditions make different negotiation behaviors appear effective. These results suggest that evaluative conclusions in this setting are shaped by the evaluation setup -- including how partial commitments are valued, which interaction protocol is used, and which game regimes are considered -- and highlight these choices as an important object of analysis for sequential negotiation, rather than treating them as a fixed background to evaluation.

\section{Related Works}
\label{sec:related_works}

Negotiation has been studied across a wide range of settings that differ in the object of negotiation (final outcomes vs.\ intermediate actions or commitments), the number of parties, the temporal structure of interaction (sequential vs.\ simultaneous), the information available to agents, and whether proposals remain non-binding or become persistent constraints. These five factors shape not only strategic behavior, but also what kinds of negotiation competence existing benchmarks can meaningfully assess. We therefore ask which negotiation regimes are already well covered by current benchmarks, and which remain under-evaluated.

\textbf{Benchmarks for outcome-level bargaining games.~}
A large fraction of negotiation research evaluates agents in \emph{outcome-level} bargaining settings, where agents negotiate over a final allocation and offers remain non-binding until acceptance. \citep{rubinstein1982perfect}'s alternating-offers model formalized this protocol and has shaped many subsequent benchmarks. The Automated Negotiating Agents Competition (ANAC), for example, established a widely used benchmark suite for bilateral multi-issue negotiation under private preference profiles, which primarily uses this alternating-offers model \citep{baarslag2010}. Outcome-level bargaining has also been instantiated in language-based settings such as Deal or No Deal \citep{lewis2017dealdealendtoendlearning} and CraigslistBargain \citep{he2018decouplingstrategygenerationnegotiation}, and more recent work extends such bargaining to multilateral LLM settings \citep{abdelnabi2024cooperationcompetitionmaliciousnessllmstakeholders}. These benchmarks have yielded important insights, but they primarily support evaluation of negotiation over final outcomes rather than settings in which intermediate commitments persist and reshape future bargaining opportunities.

\textbf{Coalition formation and coalitional bargaining.~}
Multi-party negotiation is also studied through coalition formation and coalitional bargaining, where agents negotiate over coalition structure and payoff allocation, often through sequential proposal/accept dynamics with non-binding offers until agreement. This literature is rich in theoretical results on solution concepts and computational complexity \citep{rahwan2011, aziz2011complexitycoalitionstructuregeneration}, but, to the best of our knowledge, lacks a dominant benchmark suite for systematic empirical evaluation. Existing empirical work therefore often relies on small synthetic instances \citep{chalkiadakis2004} or domain-specific settings such as collaborative vehicle routing \citep{mak2023coalitionalbargainingreinforcementlearning}. Even in these multi-party settings, however, negotiation is typically framed around coalition or payoff outcomes rather than sequential commitments that directly modify an evolving state.

\textbf{Sequential action-level negotiation with commitments.~}
Closest to our setting are negotiation environments in which agents interact sequentially through \emph{binding commitments} whose acceptance constrains future actions. For example, \citep{zhu2025learningnegotiatevoluntarycommitment} introduce Markov Commitment Games, in which agents propose joint commitments that -- if mutually accepted -- constrain future actions, but their evaluation focuses on small mixed-motive matrix games and is primarily in service of an algorithmic contribution. Other environments capture commitment-like dynamics only within domain-specific settings. Diplomacy, for instance, is a multi-party, long-horizon game in which agents coordinate and compete through committed actions, often mediated by language \citep{meta_cicero}, and Welfare Diplomacy adapts this setting to benchmark cooperative behavior among language models \citep{mukobi2023welfarediplomacybenchmarkinglanguage}. RICE-N similarly embeds a multilateral negotiation stage over policy commitments inside a climate-economy simulator \citep{zhang2022aiglobalclimatecooperation}. These environments demonstrate the importance of commitment and long-horizon interaction, but they either entangle negotiation with domain-specific mechanics or study commitment in narrow game families, making it difficult to isolate which structural properties drive performance.

Taken together, prior work provides strong evaluation settings for outcome-level bargaining and several important domain-specific environments with commitment dynamics, but leaves a gap in the evaluation of multi-party, sequential, action-level negotiation under binding commitments. Our benchmark is designed to address this gap. Additionally, rather than treating this setting as fixed -- with the sole goal of optimizing performance within a given setup -- we study how different structural and strategic assumptions change which negotiation behaviors appear effective.

\section{Negotiation Game Benchmark}
\label{sec:preliminaries}

We now describe the benchmark and its instantiation. We begin with domain insights that motivate the game design (Sec.~\ref{sec:interviews}), then formalize the negotiation game (Sec.~\ref{sec:game_formalization}), introduce the configurable generator (Sec.~\ref{sec:game_generation}), specify the reference protocol and its assumptions (Sec.~\ref{sec:gameplay_assumptions}), and finally present the baselines and evaluation methodology (Sec.~\ref{sec:baseline},~\ref{sec:evaluation}).

\subsection{Insights from Domain Experts and Real-World Practice}
\label{sec:interviews}
Our benchmark design is driven by prior studies of frontline humanitarian negotiators and the structural realities of global climate conferences like COP30 \citep{ma2025chatgpt, depledge2005organization,yamin2004negotiationprocess}, as well as our own interviews with frontline negotiators (see App.~\ref{sec:interview}). These sources find real-world agreements rarely emerge from single-step bargaining; rather, they are built through repeated bilateral and small-group consultations.

These observations directly motivate our environment formalization: a multi-party game with sequential deal-making, binding commitments, and path-dependent state updates. In this setting, partner selection, commitment ordering, and multi-goal risk analysis become central strategic concerns. To study this regime systematically, we introduce a configurable game generator that enables controlled evaluation across key structural dimensions while retaining a clear real-world analogue. We complement these synthetic sweeps with large ``real-life negotiation games” computationally derived from climate negotiation position papers from a structured negotiation exercise. Additional details on the domain context and the document-to-game translation process are provided in Appendix~\ref{app:topfile_creation}.

\subsection{Game Formalization}
\label{sec:game_formalization}

\textbf{Players and Commitments.~}
In our game, several players negotiate over a finite number of turns $T$ to reach a joint commitment that satisfies various goals, with each player valuing the resulting multi-goal satisfaction differently.

Specifically, there are $N_P$ players $p_1,\dots,p_{N_P}s$. Each player $p_n$ has a binary commitment vector $C_n$, where $C_{nj} = 1$ indicates that $p_n$ commits to doing $j$. The joint commitment state at time $t$ is $\mathbf{C}^{(t)} = (C_1^{(t)}, \dots, C_{N_P}^{(t)})$. Commitments are binding, meaning that entries may flip $0\to 1$ but never $1\to 0$. At $t=0$, $\mathbf{C}^{(0)}$ is all zeros.

\textbf{Goal Satisfaction and Reward.~}
Players negotiate to satisfy various goals according to how much they care about each goal. In a game, there are $N_G$ goals $g_1,\dots,g_{N_G}$. Player-goal utilities are given by $\mathbf{G} \in \mathbb{R}^{N_G \times N_P}$, where $\mathbf{G}_{g,n}$ is the utility player $p_n$ assigns to goal $g$ (positive, negative, or zero). 

Each goal $g$ is associated with a set of required commitments $\mathcal{C}_g$, consisting of indices $(n, j)$ into the joint commitment state $\mathbf{C}^{(t)}$. Thus, $\mathcal{C}_g$ specifies which entries $\mathbf{C}_{nj}^{(t)}$ determine goal $g$'s satisfaction.

\textit{Goal Types (Linear vs All-or-Nothing).~} We consider linear and all-or-nothing goals. For a linear goal, satisfaction is the fraction of required commitments that have been made: $S_g(\mathbf{C}^{(t)}) = \frac{1}{|\mathcal{C}_g|} \sum_{(n,j)\in \mathcal{C}_g} \mathbb{I} \left[ \mathbf{C}_{nj}^{(t) } = 1 \right]$. For an all-or-nothing goal, satisfaction is binary: $S_g(\mathbf{C}^{(t)}) = 1$ iff $\mathbf{C}_{nj}^{(t)} = 1$ for all $(n,j) \in \mathcal{C}_g$. Otherwise, $S_g(\mathbf{C}^{(t)}) = 0$.
$S(\mathbf{C}^{(t)})$ denotes the vector that contains the satisfaction value for every goal $g$.

While a utility vector $R(\mathbf{C}^{(t)}) = \mathbf{G}^\top S(\mathbf{C}^{(t)})$ can conceptually be computed at any step $t$, rewards are only realized at the terminal step $T$. Players receive $R(\mathbf{C}^{(T)})$ with no intermediary payoffs.

\subsection{Random Game Generation}
\label{sec:game_generation}
Building on this formalization, we introduce a configurable generator that produces diverse negotiation games. By varying key structural properties, it enables controlled evaluation across multiple axes of strategic difficulty and complexity.
The controllable axes include (i) the game's scale (number of players, commitments per player, and goals), and (ii) incentive alignment (whether a game is more cooperative -- players' utilities are often similar -- or more adversarial -- players' utilities are often opposed, though all games remain mixed-motive), and (iii) payoff structure (whether utilities are balanced, positive-dominated, or negative-dominated). We further control (iv) goal complexity (commitments required per goal), (v) goal non-linearity (the fraction of all-or-nothing (binary) versus linear goals), and (vi) the preference structure, which determines how strongly utilities are correlated across goals. Finally, we optionally inject (vii) \emph{poison-pill} structures: specific all-or-nothing goal patterns that create strategically deceptive offers.

We highlight goal non-linearity as a key difficulty axis in negotiation games: binary goals demand long-horizon planning, potentially forcing agents to accept short-term costs for delayed payoffs. Conversely, linear goals offer more straightforward, incremental reward signals with each commitment. 

We provide further details on these axes, including the poison-pill construction, in Appendix~\ref{app:game_generation}. Overall, our generator produces diverse sequential negotiation games under a small set of interpretable hyperparameters, yielding benchmark instances that span a wide range of strategic difficulty.

\subsection{Gameplay Assumptions and Dynamics}
\label{sec:gameplay_assumptions}

%
%

\textbf{Information Structure (Tested Approach).~}
We provide an implementation that assumes perfect information: all players know everyone's goal preferences (with full access to $\mathbf{G}$). Players also observe the entire game state $\mathbf{C}^{(t)}$ at all times $t$ and are aware of all goal requirements $\mathcal{C}_g$. Furthermore, we assume that negotiation is sequential and bilateral: at each turn, a single pair of players negotiates and updates $\mathbf{C}^{(t)}$, so the next state $\mathbf{C}^{(t+1)}$ is fully determined by the current pair's decisions.

\textbf{Negotiation Protocol (Tested Approach).~}
In our approach, players are purely self-interested: at each decision point, player $p_n$ aims to maximize their own terminal payoff $r_n = R_{p_n}(\mathbf{C}^{(T)})$. To enable this, we define a value function $V$ that  maps the current commitment state $\mathbf{C}^{(t)}$ to the vector of expected terminal payoffs for all players. $V$ can be viewed as an oracle continuation value under our protocol; in practice, we instantiate it with an approximation $\hat{V}$. Under this protocol, finding an effective approximation $\hat{V}$ becomes the central challenge.

\textit{Game Dynamics.~}Each turn has a proposer $p_n$, decided via a round-robin order, and we ensure that each player proposes equally often, with the number of proposer turns per player set as a hyperparameter. Each turn proceeds in three phases: (i) the partner selection phase, where $p_n$ selects a partner $p_{n'}$; (ii) the offer selection phase, where $p_n$ formulates their best offer to $p_{n'}$; and (iii) the decision phase, where $p_{n'}$ either accepts or rejects $p_n$'s offer. In our experiments, partner selection is handled by a sampling-based approach, while offer construction and acceptance follow a deterministic game-theoretic procedure given $\hat{V}$. Further details are in Appendix~\ref{app:negotiation_protocol}.

\textbf{Information Structure (Alternative Approaches).~}
Because the underlying game formalization is separable from any particular information structure or negotiation protocol, our game generator can also be used to evaluate agents under richer forms of uncertainty. For example, it supports partial observability, where players do not have access to others' utilities and must estimate them throughout the negotiation. It can also support simultaneous negotiation settings, where multiple pairs negotiate in parallel, making $\mathbf{C}^{(t+1)}$ dependent on concurrent agreements and introducing strategic uncertainty about others' decisions. Other sources of uncertainty can also be incorporated.

\textbf{Negotiation Protocol (Alternative Approaches).~}
While our negotiation protocol assumes self-interested play, the game formalization readily supports alternative dynamics. For example, players could be welfare-oriented (maximizing social welfare) or fairness-oriented (maximizing minimum payoff). In such variants, partner-selection, offer-construction, and acceptance rules would optimize the chosen criterion rather than individual rationality ($V_{p_n}$). 

\textbf{Summary.~}
Our benchmark is intentionally modular: the core game formalization easily pairs with varied information structures (perfect vs. partial observability; sequential vs. simultaneous) and negotiation dynamics (self-interested vs. welfare-/fairness-oriented). Combined with the generator's control over structural properties, this yields a broad family of negotiation environments spanning diverse strategic regimes, enabling systematic algorithmic evaluation under varying behavioral and informational assumptions.

\subsection{Baselines}
\label{sec:baseline}
\label{value_approximations}

We provide a small set of basic value-function approximations, spanning both planning-based and learning-based approaches, as reference solution methods rather than exhaustive attempts at optimal performance. Planning-based $\hat{V}$ have access to the full underlying game specification (e.g., rewards for every possible state), whereas learning-based $\hat{V}$ observe only gameplay trajectories paired with terminal rewards. Further details on exact implementation can be found in Appendix~\ref{app:value_function_approximations}.

\textbf{No Negotiation.~}
From discussions with domain experts, we include a ``No Negotiation'' baseline to reflect settings in which parties fail to reach agreements. Under this baseline, players do not exchange offers and instead select commitments unilaterally from the initial empty state. Effective negotiation should therefore improve the aggregate terminal payoff relative to this unilateral reference.

\textbf{Planning-Based Approximation: Myopic Reward.~}
$\hat{V}_{\text{myopic}}$ evaluates a state only through its immediate payoff, ignoring future negotiations. It is therefore best suited to regimes where progress is incremental and local improvements are a reliable guide to final outcomes.

\textbf{Planning-Based Approximation: Upper Bound (Optimistic Completion).~}
$\hat{V}_{\text{upper}}$ assumes that positively valued goals will eventually be secured anyway. As a result, it evaluates new commitments mainly in terms of the downside they introduce, making it particularly cautious about deals that increase exposure to negatively valued goals.

\textbf{Planning-Based Approximation: Lower Bound with Credible Threats (Worst-Case Completion).~}
$\hat{V}_{\text{lower}}$ assumes that some harmful outcomes may be completed by others whenever doing so is credible. Relative to the upper bound, it is therefore more tolerant of short-term losses and places greater emphasis on securing positive progress early, which can be beneficial in regimes where temporary sacrifice is needed to unlock later gains.

Taken together, these approximations act as simple but distinct decision lenses: the myopic approximation favors immediate gains, the upper bound favors caution, and the lower bound favors progress that remains valuable even under unfavorable future play.

\textbf{Learning-Based Approximations.~}
We also include learning-based approximations to assess how well generic RL-style methods perform in this reward-sparse setting. Specifically, we test a PPO-based shared policy \cite{schulman2017proximalpolicyoptimizationalgorithms}
and an RFE-based reward-free exploration and offline planning approach \cite{zhang2024uncertaintyawarerewardfreeexplorationgeneral}, both parameterized by models shared across all actors. We evaluate these learning-based methods only on larger games, where exact planning is infeasible.

\subsection{Evaluation}
\label{sec:evaluation}

While this paper focuses on modeling negotiation as a sequential, bilateral, commitment-based game, evaluation necessarily depends on how the game is played. Because performance is shaped both by the underlying game structure and by the negotiation procedure, it is difficult to isolate these two factors completely. We therefore follow the spirit of ANAC~\cite{baarslag2010}: rather than comparing across protocol choices, we fix a particular negotiation protocol and evaluate methods within that configuration. Specifically, we use the self-interested protocol from Sec.~\ref{sec:gameplay_assumptions}.

Because payoffs are player-specific, we evaluate outcomes using aggregate terminal payoff improvement relative to the No Negotiation baseline. This provides a single summary measure of how much value a method realizes through negotiation, whether it uses an explicit value approximation $\hat{V}$ or learns the relevant decision policy directly. Concretely, for a method $A$ with terminal payoff vector $\hat{r}^A$ and the No Negotiation baseline with payoff vector $\hat{r}^{\mathrm{NN}}$, we report $\sum_{n=1}^{N_P}\left(\hat{r}^{A}_n - \hat{r}^{\mathrm{NN}}_n\right)$ that is, the total gain in terminal payoff achieved through negotiation relative to unilateral action selection. Higher values indicate that the corresponding value approximation enables more effective negotiation.

While the evaluation methodology above applies to games of any size, for sufficiently small instances we can compute the exact value function under our fixed self-interested protocol using dynamic programming over the negotiation tree. This yields the optimal payoff vector $r^\star$, allowing direct measurement of deviation from exact play. In those settings, we report the $\ell_1$ error $\sum_{n=1}^{N_P} \left|r^\star_n - \hat{r}_n\right|$. This exact analysis provides a ground-truth evaluation of approximation quality when dynamic programming remains computationally feasible.


\section{Benchmark-Produced Insights on Negotiation Game Regimes}
\label{sec:methods}

Having defined the benchmark, we now use it to generate a range of negotiation regimes and study how different assumptions about future play shape which negotiation behaviors appear effective. We analyze performance through a small set of value-function approximations that encode different ways of valuing partial commitments. Rather than treating these approximations only as decision heuristics, we use them as evaluative lenses that reveal how different assumptions make different negotiation behaviors appear effective across regimes. By sweeping key structural properties of the game, we find that no single lens dominates across regimes and use these patterns to better understand evaluation in long-horizon negotiation under binding commitments.

\begin{figure}[t]
    \centering
    \includegraphics[width=\textwidth]{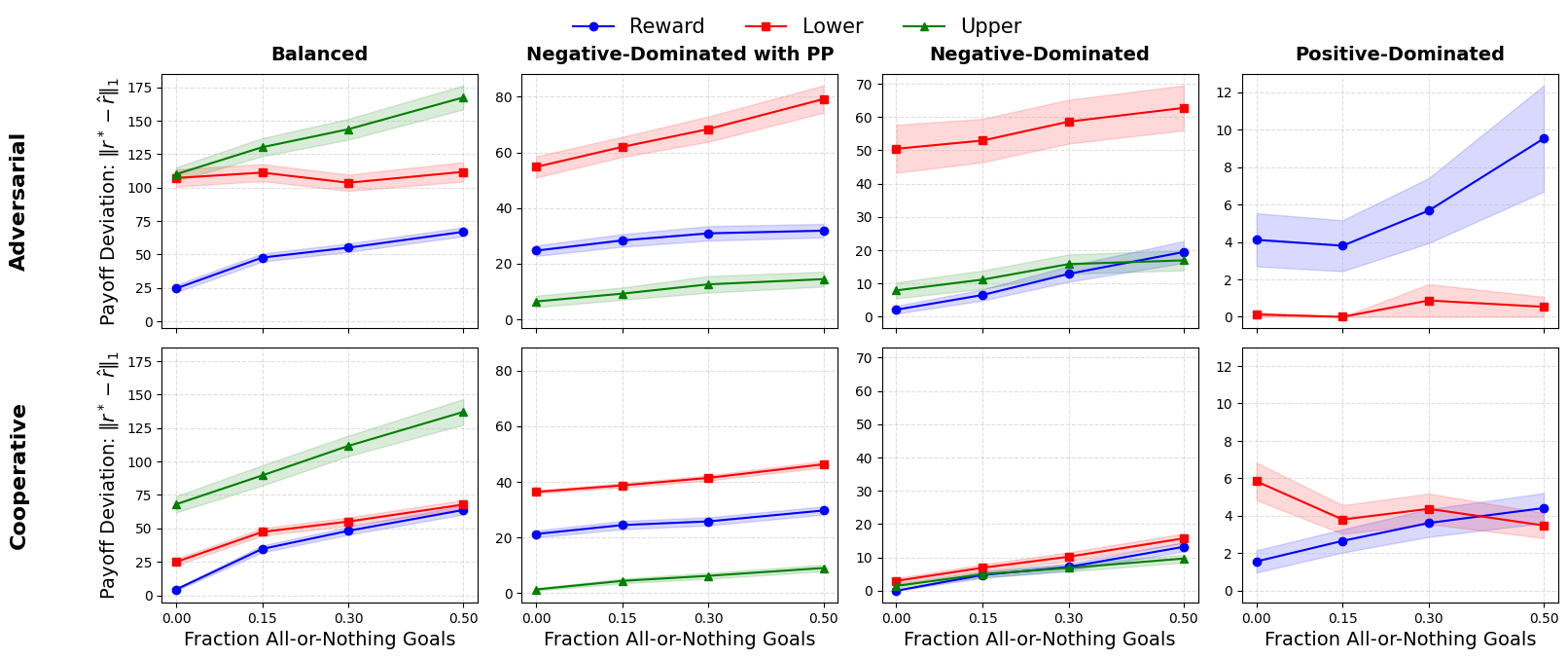}
    \caption{
    Algorithm performance on small games, measured by $\ell_1$ error to the exact optimal self-interested payoff (lower is better; columns use different y-scales). Points show mean $\pm$ standard error over 50 random games. Columns show payoff regimes (hardest to easiest), rows show incentive alignment, and the x-axis shows the fraction of all-or-nothing goals. Myopic Reward performs best in balanced games, the Upper bound in negative-dominated regimes (especially with Poison Pill traps), and the Lower bound in positive-dominated games. We omit the Upper bound in the positive-dominated regime because its errors would dominate the scale. Error generally increases with non-linearity and is higher in adversarial settings.
    }
    \label{fig:all_synthetic}
\end{figure}

\begin{table}[htbp]
    \centering
    \renewcommand{\arraystretch}{1.2}
    \begin{tabular}{@{}lcccc@{}}
        \toprule
        & \multicolumn{4}{c}{\textbf{Game Regime}} \\
        \cmidrule(l){2-5}
        \textbf{Method} & \textbf{Balanced} & \textbf{Negative-Dom. w/ PP} & \textbf{Negative-Dom.} & \textbf{Positive-Dom.} \\
        \midrule

        \multicolumn{5}{@{}l}{\textit{Adversarial}} \\
        Reward & $\mathbf{14.45 \pm 1.25}$ & $-12.81 \pm 2.65$ & $5.69 \pm 3.07$ & $46.85 \pm 1.01$ \\
        Lower  & $-34.09 \pm 5.22$ & $-196.56 \pm 14.32$ & $-185.82 \pm 10.21$ & $\mathbf{53.72 \pm 0.81}$ \\
        Upper  & $-14.48 \pm 13.09$ & $\mathbf{11.55 \pm 3.90}$ & $\mathbf{12.39 \pm 4.64}$ & $-260.52 \pm 239.92$ \\
        \addlinespace
        
        \multicolumn{5}{@{}l}{\textit{Cooperative}} \\
        Reward & $\mathbf{9.05 \pm 4.23}$ & $-3.96 \pm 2.31$ & $\mathbf{4.43 \pm 2.25}$ & $\mathbf{3.05 \pm 1.52}$ \\
        Lower  & $-13.51 \pm 4.46$ & $-24.72 \pm 3.12$ & $-2.93 \pm 2.31$ & $-3.93 \pm 1.55$ \\
        Upper  & $-49.53 \pm 29.82$ & $\mathbf{2.47 \pm 0.99}$ & $3.09 \pm 1.63$ & $-252.63 \pm 216.04$ \\
        
        \bottomrule
    \end{tabular}
    \caption{
    Aggregate improvement in terminal payoff relative to the No Negotiation baseline across payoff regimes and incentive alignments. Higher is better. For compactness, values are averaged over all fractions of binary goals and reported as mean $\pm$ standard error. Bold entries denote the best- performing method within each regime–alignment setting. More detailed results are in App.~\ref{app:results}.
    }
    \label{tab:regime_diffs}
\end{table}

\subsection{Evaluation Across Synthetic and Document-Grounded Regimes}
We evaluate the benchmark in three complementary settings. First, in small synthetic games, exact dynamic programming is tractable, allowing direct comparison to optimal self-interested play. Second, in larger synthetic games, exact planning becomes infeasible, so we evaluate methods by their improvement in aggregate terminal payoff relative to the No Negotiation baseline. Third, we study document-grounded games derived from real negotiation artifacts, which test whether the patterns observed in controlled sweeps persist in larger and more realistic instances.

Across these settings, we vary key structural properties of the game, including incentive alignment, payoff distribution, goal complexity, utility correlation, and the fraction of all-or-nothing goals. This enables us to treat the benchmark not just as a source of instances, but as an instrument for probing how evaluative conclusions change across negotiation regimes. Further details can be found in Appendix~\ref{app:benchmark_sweeps}.
Unless otherwise noted, synthetic games contain 10 players, 15 goals, and an agreement budget of four actions per turn. We generate 50 random instances for each combination of structural properties. Small generator games use 1-2 commitments per player, which keeps exact evaluation tractable. Large generator games use 5 commitments per player, making exact planning infeasible. Real-life negotiation games are document-grounded instances derived from climate negotiation materials and are substantially larger. 

This setup lets us compare two forms of evaluation. In small games, exact value computation provides a ground-truth measure of approximation quality. In larger games, exact evaluation is unavailable, so we instead ask whether a method improves realized outcomes relative to a unilateral no-negotiation reference point. Taken together, these settings allow us to examine not only algorithm behavior, but also how exact comparison to optimal play on small games relates to comparative evaluation against a no-negotiation baseline on larger ones.

\subsubsection{Small Generator Games}
\label{sec:small_results}

Figure~\ref{fig:all_synthetic} shows that approximation quality depends strongly on the structural regime of the game. Balanced games are consistently hard: they yield the largest deviations from exact optimal play, indicating that mixed-sign incentives make continuation values especially difficult to estimate. Positive-dominated games are easiest, while negative-dominated games fall in between; injecting poison-pill structures further increases difficulty. Across all payoff structures, adversarial games are harder than cooperative ones, and increasing the fraction of all-or-nothing goals generally increases error.

Just as importantly, the best-performing baseline changes across regimes. In balanced games, the Myopic Reward approximation achieves the lowest error. In negative-dominated games, the Upper lens performs best overall, especially under poison-pill constructions. In positive-dominated games, the Lower lens performs best and remains comparatively robust as non-linearity increases.

These results matter methodologically because they show that there is no regime-independent good baseline in this setting. The benchmark therefore exposes a core evaluation risk: conclusions about negotiation competence can change depending on which assumptions are built into the value model used to assess partial commitments.

\subsubsection{Large Generator Games}
\label{sec:large_results}

\begin{wraptable}[13]{r}{0.40\linewidth}
    \vspace{-15pt}
    \centering
    \renewcommand{\arraystretch}{1.15}
    \begin{tabular}{@{}lc@{}}
        \toprule
        \textbf{Method} & \textbf{Diff. from Baseline} \\
        \midrule
        Reward & $\hphantom{-}\mathbf{7.60 \pm 5.71}$ \\
        Lower  & $-0.98 \pm 7.97$ \\
        Upper  & $-9.54 \pm 12.83$ \\
        \midrule
        RFE (learning)   & $-44.65 \pm 26.56$ \\
        PPO (learning)   & $-46.00 \pm 26.38$ \\
        \bottomrule
    \end{tabular}
    \caption{
    \textbf{Large Generator Games with PPO and RFE.} Aggregate improvement in terminal payoff relative to the No Negotiation baseline (mean $\pm$ standard error). Higher is better.
    }
    \label{tab:learning_results}
\end{wraptable}

We next test whether the patterns observed under exact evaluation persist once exact planning becomes infeasible. Here we use a comparative outcome metric: aggregate improvement in terminal payoff relative to the No Negotiation baseline.
Table~\ref{tab:regime_diffs} shows that the regime-level patterns largely carry over. In adversarial settings, the same structure reappears: the Upper lens is strongest in negative-dominated games, especially with poison pills; Myopic Reward performs best in balanced regimes; and the Lower lens is strongest in positive-dominated games. In more cooperative settings, these differences narrow, suggesting that when incentives are more aligned, the choice of evaluative lens matters less because continuation uncertainty is less strategically consequential.

We also evaluate two learning-based approximations in this large-game regime. Specifically, we train a PPO-based shared policy and an RFE-based reward-free exploration and offline planning approach on balanced, adversarial games. As shown in Table~\ref{tab:learning_results}, both learning-based methods underperform the simple planning-based lenses, with PPO and RFE falling well below the No Negotiation baseline. This suggests that, in our reward-sparse setting, generic learning-based approaches struggle to recover effective negotiation behavior at this scale. Additional training details are provided in Appendix~\ref{app:learning_details}.

This result is important for the benchmark’s evaluative role. It suggests that the simpler comparative metric used in large games preserves the main qualitative conclusions observed under exact evaluation on small games. In other words, the benchmark supports a useful bridge between settings where exact evaluation is possible and settings where only surrogate comparative evaluation is tractable.

\begin{wraptable}[13]{r}{0.36\linewidth}
    \vspace{-15pt}
    \centering
    \renewcommand{\arraystretch}{1.15}
    \begin{tabular}{@{}lc@{}}
        \toprule
        \textbf{Method} & \textbf{Diff. from Baseline ($\Delta$)} \\
        \midrule
        Lower          & $\hphantom{-}\mathbf{26.61} \pm \mathbf{9.82}$ \\
        Reward         & $\hphantom{-}18.90 \pm 7.57$ \\
        Upper          & $-82.67 \pm 51.36$ \\       
        \bottomrule
    \end{tabular}
    \caption{
    \textbf{Real-Life Negotiation Games.} Aggregate improvement in terminal payoff relative to the No Negotiation baseline on five real-world negotiation games (mean $\pm$ standard error). Higher is better.
    }
    \label{tab:topfile_aggregate}
\end{wraptable}

\subsubsection{Real-Life Negotiation Games}
\label{sec:real_world_results}

The climate negotiation games derived from the position papers (details in App.~\ref{app:topfile_creation}) are too large for exact dynamic programming, so we again compare methods to No Negotiation. These games are predominantly positive-dominated, adversarial, and largely linear, though they remain strategically challenging because of their scale and mixed motives. Table~\ref{tab:topfile_aggregate} shows that the Lower lens achieves the highest aggregate improvement, while the Myopic Reward method also performs well. The Upper lens underperforms, reflecting excessive conservatism in a regime where beneficial commitments are common. Taken together, these results show that the insights from our benchmark's controlled synthetic regimes meaningfully connect to real instances.

\subsection{Interpreting Regime-Specific Performance Across Value Approximation Lenses}
\label{sec:implications}

The regime-specific winners illustrate the benchmark's value as an evaluation tool: by varying game structure, it reveals when different assumptions about partial commitments change which negotiation behaviors appear effective. These patterns reflect each approximation's inductive bias: Myopic Reward values states by immediate payoff; Upper treats positive goals as eventually realized anyway and therefore penalizes added negative exposure; Lower treats some harmful outcomes as unavoidable and therefore favors states that remain robust while securing positive progress.

In \emph{balanced} games, players face mixed-sign utilities across many goals -- creating strategic conflict -- so whether a commitment is ``good'' depends heavily on future trades. Consistently optimistic (Upper) or threat-focused (Lower) strategies can therefore be brittle: Upper may reject useful stepping-stone deals, while Lower may accept sacrifices that are not repaid later. Myopic Reward makes fewer assumptions about how the game will resolve, which is safer when upside and downside are both plausible. This conflict also aligns with the higher L1 errors in small balanced instances.

In \emph{negative-dominated} games, where avoiding penalties dominates outcomes, the Upper approximation lens performs well by behaving conservatively with respect to negative exposure. This advantage becomes most pronounced when we inject poison pill structures: offers that look locally attractive but trigger large downstream penalties. Upper’s strict filtering helps avoid these traps, whereas the myopic baseline is more likely to accept them based on immediate payoff signals.

In \emph{positive-dominated} games, most goals yield positive utility to players, so negotiation is less about avoiding penalties and more about reliably completing beneficial goals under uncertainty about follow-through. The Lower lens performs well here, as it assumes that opponents will realize credible threats, and thus prioritizes states that remain safe under unfavorable continuations. It therefore emphasizes locking in positive progress wherever possible. This regime also shows lower L1 error on small games, consistent with reduced strategic conflict when most goals are positively valued.

More broadly, these results show why a controlled benchmark is useful for this setting. Evaluation depends not only on the game instance, but also on how the downstream consequences of partial commitments are modeled. By making these assumptions explicit and varying the regimes in which they are tested, the benchmark helps identify when a negotiation method is genuinely robust and when it is merely well matched to a particular evaluative lens.

\section{Algorithmic Development Opportunities Exposed by our Benchmark}

Our benchmark generates diverse, non-trivial games and reveals regime-dependent advantages of different value-approximation lenses, with no single lens dominating across the instance space. This suggests two research directions for robust negotiation agents: (i) learning substantially more accurate state-value estimates that generalize across regimes, and (ii) developing adaptive methods that infer the regime online and select evaluation strategies accordingly.

A second direction concerns the drivers of difficulty. The small-game sweeps indicate that the hardest instances are those with stronger mixed-motive conflict, and that difficulty increases as all-or-nothing goals become more prevalent. Both factors increase the strategic weight of intermediate commitments: partial progress is hard to value reliably under all-or-nothing goals, and stronger conflict means many commitments are not jointly beneficial, increasing path dependence because commitments can sharply constrain which future agreements remain feasible. Algorithmic development is needed to reason strategically over long horizons under binding commitments with only terminal rewards.

Finally, our insights in Sec.~\ref{sec:methods} are all derived from a protocol that adopts perfect information and sequential bilateral negotiation within a multi-party game -- chosen to isolate the effect of state valuation. However, our benchmark supports many different protocols. Given the challenges in this relatively simple protocol, we expect algorithmic development to be needed for more complex protocols that involve partial observability (e.g., unknown opponent utilities requiring inference during play) and richer multi-party dynamics (e.g., multiple negotiations occurring concurrently). 

\section{Conclusion}

Motivated by real negotiation practice, we introduce a benchmark and evaluation framework for sequential multi-party negotiation with binding commitments and terminal-only rewards. We provide (i) a controllable game generator that sweeps key structural properties such as scale, incentive alignment, payoff structure, and goal non-linearity, (ii) a reference turn-taking protocol, and (iii) an evaluation suite that supports exact dynamic-programming evaluation on small games and comparative evaluation on larger games. Intentionally modular, the benchmark supports alternative negotiation protocols and information structures, such as partial observability.

We also provide several baseline solvers. Across both synthetic sweeps and real climate negotiation instances, no single approach dominates. Difficulty rises with mixed-motive conflict and with all-or-nothing goals, and different payoff structures favor different ways of valuing partial commitments. These results suggest that conclusions about which negotiation behaviors appear effective depend not only on the game instances being studied, but also on the assumptions built into the evaluation setup. While such methods may have largely beneficial applications (e.g., supporting decision-making in high-stakes scenarios), we note that they could also be misused in adversarial contexts (e.g., for strategic manipulation). Taken together, our benchmark and evaluation framework provides insights on where algorithmic development is needed to support complex negotiation settings.

\bibliography{bibliography}
\bibliographystyle{abbrv}
\newpage

\appendix

\section{Interview with Domain Experts}
\label{sec:interview}

Our negotiation game benchmark design is primarily informed by prior interview-based work with frontline humanitarian negotiatiors \citep{ma2025chatgpt}, which identifies negotiation preparation as a multi-step, process-oriented activity involving context analysis, compromise ideation, risk analysis, and knowledge sharing; it also emphasizes that negotiators prefer AI support that helps structure and explore options while preserving human judgment and collaboration. These findings directly motivate several operational assumptions in our environment design: sequential deal-making, the strategic importance of partner selection and commitment ordering, explicit reasoning about trade-offs and risks across multiple goals, and the need to recover \textit{sequences} of self-interested decisions (not only final outcomes) for preparation tasks such as stakeholder analysis and actor-influence planning \cite{CCHN}.

COP30 -- the 30th Conference of the Parties (COP) to the United Nations Framework Convention on Climate Change (UNFCCC), hosted in Belém, Brazil -- provides an intuitive process template for our abstraction. During the conference, agreement was not reached in a single step, but emerged through repeated bilateral and small-group consultations that incrementally added progress (through commitments) towards multiple objectives. Formal sessions (e.g., plenaries and contact groups) were interleaved with a large number of informal consultations and Presidency-/ministerial-led discussions that brokered compromises over time. This motivates modeling negotiation as sequential deal-making over binding commitments, where partner selection and commitment ordering matter, and where parties must explicitly reason about multi-goal trade-offs and risks as the agreement gradually grows \citep{unfccc_road_to_belem,enb_cop30_10nov2025,enb_cop30_13nov2025}.

Motivated by these observations, we formalize this process as a sequential, multi-party negotiation environment with repeated bilateral interactions, binding commitments, and path-dependent state updates, while retaining core process features emphasized in studies of UNFCCC negotiation practice \citep{depledge2005organization,yamin2004negotiationprocess}.

Because this negotiation regime remains underrepresented in literature (see Sec.~\ref{sec:related_works}), we introduce a benchmark in the form of a configurable game generator that enables controlled sweeps over key structural dimensions, supporting the rigorous evaluation of algorithms and models in settings where long-horizon consequences and path dependence are central. 


We further create large benchmark instances from real negotiation artifacts by converting position papers from a climate negotiation exercise into machine-readable negotiation games, which we call “real-life negotiation games.” In this exercise, participants role-play different countries in a multilateral climate negotiation; accordingly, we treat these documents as realistic role-play negotiation inputs rather than literal state policy documents. Additional details on the source and processing of these documents are provided in Appendix \ref{app:topfile_creation}. The underlying position papers are included in the supplementary materials.

\section{Real-Life Negotiation Games Creation}
\label{app:topfile_creation}

We generated each topfile from the set of position papers submitted by the negotiating parties using a large language model pipeline (implemented with GPT-5.2) with intermediate outputs.

First, we treated each position paper as one party’s statement of preferences and commitments. From each paper, the model extracted two kinds of issue-level content: the outcomes the party most wanted to secure, and the outcomes it most wanted to avoid. At this stage, the extraction was designed to capture underlying negotiation issues rather than concrete commitments, a framework used by frontline negotiators (the redline/ bottomline tool)~\cite{CCHN}. For example, the system aimed to extract goals such as control over a port, protection of sovereignty, or preferred security terms, while excluding concrete instruments such as signing a compact, funding a package, or deploying a mission.

Second, because the same underlying issue can appear in different papers with different wording, we canonicalized the extracted issue statements into a shared set of distinct negotiation goals. This canonicalization step considered all extracted goals together and rewrote them into common scenario-level labels. Each canonical goal was retained only when it could be tied back to explicit supporting evidence from one or more extracted source statements. This produced a common goal set that could be used across all parties.

Third, once this shared goal set was established, we estimated each party’s stance toward each goal. For every party-goal pair, the model assigned a discrete score indicating whether the party strongly supported the goal, moderately supported it, was neutral, moderately opposed it, or strongly opposed it.

Fourth, we extracted each party’s concrete actions from its position paper. Here the task was the inverse of goal extraction: instead of recovering desired outcomes, the model identified executable commitments or moves that the party could take, such as granting access, financing a package, issuing a pledge, or joining an agreement. Repeated extractions were then reconciled into a deduplicated canonical action list for each party, collapsing near-duplicate phrasings into a single action representation. These action lists defined the feasible action set for each party.

Fifth, we mapped actions onto goals. For each party, the model evaluated which of that party’s concrete actions would materially contribute to satisfying each canonical goal. This step linked the actions to the preferences and produced the action-to-goal structure used in the game representation.

Sixth, we classified which goals should be treated as binary rather than gradual. Binary goals were those that were best represented as achieved or not achieved, rather than partially satisfied along a continuum.

To improve robustness, the main semantic extraction stages were run multiple times and reconciled using consensus procedures. Goal extraction and action extraction used repeated model runs followed by clustering and aggregation to reduce sensitivity to wording variation. Goal scoring, action-to-goal mapping, and binary classification also used repeated runs with majority-style aggregation. This design reduced the influence of any single stochastic model output.

Finally, we assembled the topfile by combining: the number of parties, the number of actions available to each party, the party-by-goal preference scores, the mapping from goals to supporting actions, the set of binary goals, and the user-specified procedural parameters of the game, such as action cost and bargaining order. 

To assess stability, we generated five topfiles from the same set of position papers using the current pipeline and compared outputs across all ten run pairs. Using a symmetric best-match cosine over canonical goal labels, the mean goal similarity was 0.791. Using a best-match cosine measure within actor across runs, the mean action similarity was 0.960.

\section{Random Game Generation}
\label{app:game_generation}

In Section~\ref{sec:game_generation}, we state seven controllable game properties. While scale (property 1) is directly specifiable, we detail the remaining axes here.

\textbf{Property: Correlation of Goal Utilities.~}
The goal-utility matrix is generated with dependence on the hyperparameter $d$, which represents the number of hidden latent factors driving preferences. We sample a $d$-dimensional latent vector for every goal $g$, $v_g \in \mathbb{R}^d$, and for every player $p$, $u_p \in \mathbb{R}^d$, from a Gaussian distribution and calculate goal-utility values like this: $
{G}_{g,p} = v_g^\top u_p \;+\; \epsilon_{g,p}\;$, where  $\epsilon_{g,p} \sim \mathcal{N}(0, \sigma^2)
$ adds random noise. We rescale the matrix ${G}$ to a target integer range $[g_{\min}, g_{\max}]$ and round to integers; in our main experiments we use $g_{\min}=-30$ and $g_{\max}=30$. 

The degree of utility correlation across goals is directly controlled by the number of latent factors $d$: a smaller $d$ induces stronger shared structure across goals (utilities lie in a lower-dimensional subspace, increasing correlations), while a larger $d$ allows preferences to depend on many independent factors, yielding more uncorrelated goal utilities.

\textbf{Property: More Cooperative vs More Adversarial Games.~}
To control how aligned player's preferences are, we vary the distribution of player vectors $u_p$.  If the game is specified to be more cooperative, we sample $u_p$ from a Gaussian with positive mean (biasing utilities toward similar signs across players). In more adversarial games, we sample with mean zero, producing more mixed signs and hence more frequent preference conflicts.

\textbf{Property: Payoff Structure.~}
We control whether a game's utilities are balanced, positive-dominated, or negative-dominated by shifting the bounds of the target utility range $[g_{\min}, g_{\max}]$ (we use $[-20, 20]$ for balanced, $[-8, 30]$ for positive-dominated, and $[-30, 8]$ for negative-dominated).

\textbf{Property: Linear vs All-or-Nothing Goals.~}
The generator specifies the fraction of all-or-nothing goals, which are assigned uniformly at random. This is a key difficulty axis: all-or-nothing goals demand long-horizon planning, which may be forcing agents to accept short-term tradeoffs for delayed payoffs. Conversely, linear goals offer more straightforward, incremental reward signals with each commitment.

\textbf{Property: Complexity of Goals.~}
One can also specify the hyperparameter $\alpha$, which specifies how many commitments are generally required for a goal. We use a Zeta distribution to sample the number of commitments per goal, $c_g \sim \text{Zeta}(\alpha)$, where $\alpha$ controls the heaviness of the tails, leading to more commitments needed per goal. We upper-bound $c_g$ so that it makes sense given the other parameters of the game. We also ensure that $c_g \geq 1$ for linear goals and $c_g\ge 2$  for all-or-nothing goals (as they would otherwise be trivial). We then sample $c_g$ distinct $(n, j)$ player-commitment indices uniformly (without replacement) from the set of all possible indices to form the set  $\mathcal{C}_g$ of required player-commitments for goal $g$.

\textbf{Property: Injected Poison-Pill Structures.~}
To introduce an additional and controlled source of difficulty, we optionally inject a specific binary-goal structure that we call a \emph{poison pill}. Because all-or-nothing goals are otherwise assigned uniformly at random, this mechanism allows us to deliberately create games containing locally attractive but strategically harmful bundled offers. In particular, a poison pill creates situations in which one player can propose a deal that combines a mutually beneficial commitment with an additional commitment that is costly to the partner, even though the beneficial part could have been secured later without accepting the harmful one. This makes deal evaluation more demanding, since agents must reason not only about immediate gains but also about whether a proposed bundle contains avoidable downstream costs.

Concretely, consider a poison-pill interaction between proposer $p_n$ and partner $p_{n'}$. There are two goals: a mutually beneficial ``bait'' goal requiring joint action from both players, and a ``poison pill'' goal that benefits $p_n$ but harms $p_{n'}$ and requires action only from $p_{n'}$:
\[
\begin{aligned}
\mathcal{C}_{g_{\text{bait}}} &= \{(p_n,0),(p_{n'},0)\}, 
&\quad \mathbf{G}_{g_{\text{bait}},p_n} &= 30,\;
\mathbf{G}_{g_{\text{bait}},p_{n'}} = 30, \\
\mathcal{C}_{g_{\text{poison}}} &= \{(p_{n'},1)\}, 
&\quad \mathbf{G}_{g_{\text{poison}},p_n} &= 10,\;
\mathbf{G}_{g_{\text{poison}},p_{n'}} = -25.
\end{aligned}
\]

If $p_n$ proposes completing both goals simultaneously, the deal appears immediately acceptable to a myopic evaluator because the net payoff to $p_{n'}$ is still positive ($30 - 25 = 5$). However, this is strategically suboptimal: $p_{n'}$ could reject the bundled offer and still realize the bait goal later, without taking on the poison pill, since the bait is jointly beneficial to both players. By contrast, stricter value approximations can recognize this structure and reject the offer. Injecting poison pills therefore guarantees an additional source of difficulty beyond the random assignment of binary goals, creating environments in which successful negotiation requires distinguishing genuinely beneficial cooperation from bundled commitments that exploit short-term reasoning.

\section{Negotiation Protocol}
\label{app:negotiation_protocol}

In Section~\ref{sec:gameplay_assumptions}, we introduced our negotiation protocol. We elaborate further on our assumptions here.

\textit{Game Dynamics.~}Each turn has a proposer $p_n$, decided via a round-robin order, and we ensure that each player proposes equally often, with the number of proposer turns per player set as a hyperparameter. Each turn proceeds in three phases: (i) the partner selection phase, where $p_n$ selects a partner $p_{n'}$; (ii) the offer selection phase, where $p_n$ formulates their best offer to $p_{n'}$; and (iii) the decision phase, where $p_{n'}$ either accepts or rejects $p_n$'s offer. The entire game is described by Algorithm~\ref{alg:solvegame}.

\textit{Partner Selection.~} The proposer $p_n$ seeks to select the partner $p_{n'}$ that maximizes their expected terminal payoff given the current state $\mathbf{C}^{(t)}$. We use Monte Carlo Tree Search (MCTS) with $\hat{V}$ as a leaf evaluator for this selection, reducing the brittleness of greedy $\arg\max \hat{V}$ choices (see Algorithm~\ref{alg:mcts}). We treat offer construction and acceptance decisions as a downstream subroutine scored by $\hat{V}$.

\textit{Offer Selection and Decision Phase.~}
After selecting partner $p_{n'}$, proposer $p_n$ searches for a feasible joint commitment update $\Delta C_{n,n'}^{(t)} = (\Delta C_n^{(t)}, \Delta C_{n'}^{(t)})$ -- where $\Delta C_n^{(t)}$ indicates the new commitments of player $p_n$ at turn $t$. Offers must be: (i) \emph{binding} (flipping 0s to 1s only) and (ii) add at most $k$ new commitments per player per turn, reflecting that real-world interactions typically only lead to a small number of agreements (according to domain experts). More formally, the proposer $p_n$ and their partner $p_{n'}$ enter an ultimatum bargaining game where $p_n$ solves:
\begin{equation}
\label{eq: offer_selection}
\Delta C_{n,n'}^{(t),\;\star} = 
\argmax_{\Delta C_{n,n'}^{(t)}\in \mathcal{F}(\mathbf{C}^{(t)})}\;\; V_{p_n}\left(\mathbf{C}^{(t)} \oplus \Delta C_{n,n'}^{(t)}\right) 
\quad \quad \text{s.t.} \quad
V_{p_{n'}}\left(\mathbf{C}^{(t)} \oplus \Delta C_{n,n'}^{(t)}\right) \geq V_{p_{n'}}\left(\mathbf{C}^{(t)}\right)
\end{equation}
where $V$ maps the commitment state $\mathbf{C}^{(t)}$ to the expected terminal payoff vector, assuming all players act optimally to maximize their individual payoffs in all subsequent turns, $\mathcal{F}(\mathbf{C}^{(t)})$ denotes the set of monotone (flipping 0s to 1s only) feasible updates satisfying $\lVert \Delta C_n^{(t)} \rVert_0 \leq k$ and $\lVert \Delta C_{n'}^{(t)} \rVert_0 \leq k$, and where $\oplus$ denotes $\mathbf{C}^{(t)}$ updated with the commitments specified by $\Delta C_{n,n'}^{(t)}$. If a solution exists, $p_{n'}$ accepts (as their value weakly improves), yielding $\mathbf{C}^{(t+1)} = \mathbf{C}^{(t)} \oplus \Delta C_{n,n'}^{(t), \;\star}$. If there is no solution -- meaning that there is no offer that would be acceptable for the partner -- the partner rejects and the game proceeds to the next turn: $\mathbf{C}^{(t + 1)} = \mathbf{C}^{(t)}$.

\begin{algorithm}[H]
\caption{\textsc{SolveGame}}
\label{alg:solvegame}
\begin{algorithmic}[1]
    \Statex \textbf{Input:} Value approximation $\widehat{V}^{*}$, horizon $T$
    \Statex \textbf{Output:} Terminal agreement matrix $\GameState^{(t)}$ and history $\{(r_t,\mathbf{a}_t^{*})\}_{t=0}^{T-1}$
    \Statex \hrulefill
    
    \State $t \gets 0$
    \State $s_0 \gets \GameState_0$ \Comment{Initialize agreement matrix}
    
    \While{$t < T$}
        \State $r_t \gets \textsc{Mcts}(s_t, \widehat{V}^{*})$ \Comment{Select partner via search}
        \State $\mathbf{a}_t^{*} \gets \textsc{GetBestOffer}(s_t, r_t, \widehat{V}^{*})$ \Comment{Solve Eq.~\ref{eq: offer_selection}}
        
        \If{$\mathbf{a}_t^{*} \neq \varnothing$}
            \State $s_{t+1} \gets \textsc{PlayDeal}(s_t, \mathbf{a}_t^{*}, r_t)$ \Comment{$\GameState^{(t+1)}\leftarrow \GameState^{(t)}\oplus \mathbf{a}_t^{*}$}
        \Else
            \State $s_{t+1} \gets \textsc{RejectDeal}(s_t)$ \Comment{No change: $\GameState^{(t+1)}=\GameState^{(t)}$}
        \EndIf
        
        \State $t \gets t + 1$
    \EndWhile
    
    \State \Return $\GameState^{(t)}$, $\{(r_t,\mathbf{a}_t^{*})\}_{t=0}^{T-1}$
\end{algorithmic}
\end{algorithm}

\begin{algorithm}[H]
\caption{\textsc{MCTS}}
\label{alg:mcts}
\begin{algorithmic}[1]
    \Statex \textbf{Input:} Root state $s_{\text{root}}$, simulations $N_{\text{sim}}$, exploration constant $c_{\textsc{ucb}}$, value approximation $\widehat{V}^{*}$
    \Statex \textbf{Output:} Selected partner $r^{*}$
    \Statex \hrulefill
    
    \State $root \gets \textsc{Node}(s_{\text{root}})$
    
    \For{$n = 1$ \textbf{to} $N_{\text{sim}}$}
        \State $s \gets s_{\text{root}}.\textsc{Clone}()$
        \State $node \gets root$
        
        \Comment{\textbf{1. Selection}}
        \While{$node$ is fully expanded \textbf{and not} $s.\textsc{IsTerminal}()$}
            \State $p \gets s.\textsc{Proposer}()$
            \State $r \gets \arg\max_{j \in node.children}\ \textsc{UcbValue}(node.child(j), node.N, p, c_{\textsc{ucb}})$
            \State $\mathbf{a} \gets \textsc{GetBestOffer}(s, r, \widehat{V}^{*})$ \Comment{Solve Eq.~\ref{eq: offer_selection}}
            
            \If{$\mathbf{a} \neq \varnothing$}
                \State $s.\textsc{PlayDeal}(\mathbf{a}, r)$
            \Else
                \State $s.\textsc{RejectDeal}()$
            \EndIf
            
            \State $node \gets node.child(r)$
        \EndWhile
        
        \Comment{\textbf{2. Expansion}}
        \If{\textbf{not} $s.\textsc{IsTerminal}()$}
            \State $r \gets node.untried\_partners.\textsc{Pop}()$
            \State $\mathbf{a} \gets \textsc{GetBestOffer}(s, r, \widehat{V}^{*})$ \Comment{Solve Eq.~\ref{eq: offer_selection}}
            
            \If{$\mathbf{a} \neq \varnothing$}
                \State $s.\textsc{PlayDeal}(\mathbf{a}, r)$
            \Else
                \State $s.\textsc{RejectDeal}()$
            \EndIf
            
            \State $new\_node \gets \textsc{Node}(parent=node, action=r)$
            \State $node.child(r) \gets new\_node$
            \State $node \gets new\_node$
        \EndIf
        
        \Comment{\textbf{3. Simulation / evaluation}}
        \State $R \gets s.\textsc{GetPayoffVector}()$
        
        \Comment{\textbf{4. Backpropagation}}
        \While{$node \neq \textsc{Null}$}
            \State $node.N \gets node.N + 1$
            \State $node.W \gets node.W + R$ \Comment{$R\in\mathbb{R}^{\Players}$}
            \State $node \gets node.parent$
        \EndWhile
    \EndFor
    
    \State \Return $\arg\max_{r \in root.children}\ root.child(r).N$
\end{algorithmic}
\end{algorithm}

\section{Value Function Approximations}
\label{app:value_function_approximations}

As described in Sec.~\ref{sec:gameplay_assumptions}, our negotiation protocol relies on a value function $V$ mapping the current commitment state $\mathbf{C}^{(t)}$ to the expected terminal payoff vector. While $V$ can be computed via dynamic programming in small games (Algorithm~\ref{alg:exact_solver}), this quickly becomes intractable as the game size grows. We therefore consider three simple approximations $\hat{V}$, each encoding a different assumption about how the rest of the game may unfold. In the following sections, we use these approximations both as decision heuristics and as diagnostic lenses in our regime sweeps.

\textbf{Value Approximation 1: Myopic Reward.~} 
This approximation ignores future negotiations, it evaluates a joint commitment only by its immediate impact on the payoff vector: $\hat{V}^{\text{reward}}(C^t) \;=\; R(C^t)$.
Intuitively, this means $\hat{V}^{\text{reward}}$ tends to work best when progress is incremental -- e.g., in games with mostly linear goals.

\textbf{Value Approximation 2: Upper-bound (Optimistic Completion).~} 
Given a player $p_n$, this approximation assumes that player $p_n$ will secure all remaining rewards from \textit{positively valued goals}:
\begin{equation*}
\resizebox{\linewidth}{!}{$\displaystyle
\hat{V}^{\text{upper}}_{p_n}(\mathbf{C}^{(t)}) 
= R_{p_n}(\mathbf{C}^{(t)}) + \sum_{g:\mathbf{G}_{g,n}>0} \bigl(1-S_g(\mathbf{C}^{(t)})\bigr)\mathbf{G}_{g,n} 
= \sum_{g:\mathbf{G}_{g,n}>0} \mathbf{G}_{g,n} + \sum_{g:\mathbf{G}_{g,n}<0} S_g(\mathbf{C}^{(t)})\mathbf{G}_{g,n}
$}
\end{equation*}

Accordingly, new commitments are not judged by how much more value they add -- as they all feed into the constant $\sum_{g:\mathbf{G}_{g,n}>0} \mathbf{G}_{g,n}$ -- but by how much value they could take away: $\sum_{g:\mathbf{G}_{g,n}<0} S_g(\mathbf{C}^{(t)})\mathbf{G}_{g,n}$. This results in the player being highly cautious about deals that increase the satisfaction of goals with negative effects.

\textbf{Value Approximation 3: Lower-Bound with Credible Threats (Worst-Case Completion).~}
Given a player $p_n$, it assumes future players will complete goals that are negative for $p_n$. However, we restrict attention to credible threats $\mathcal{G}_{\textsc{CT}}(p_n)$. These are goals that $p_n$ values negatively but remain non-negative for every other required contributor. This filters out self-harming threats aimed at hurting $p_n$: \\
$
\hat{V}^{\text{lower}}_{p_n}(\mathbf{C}^{(t)})
\;=\;
R_{p_n}(\mathbf{C}^{(t)})
\;+\;
\sum\limits_{g \in \mathcal{G}_{\textsc{CT}}(p_n)}
\bigl(1-S_g(\mathbf{C}^{(t)})\bigr)\,\mathbf{G}_{g,n}
$
where $\mathcal{G}_{\textsc{CT}}$ is formally defined as:

$\mathcal{G}_{\textsc{CT}}(p_n) = \left\{ g \in \mathcal{G} \;\middle|\; \mathbf{G}_{g,n} < 0 \text{ and } \mathbf{G}_{g,m} \ge 0 \text{ for all } p_m \neq p_n \text{ with pending actions for } g \right\}$

State valuations become driven primarily by positive progress and more tolerant of immediate negative exposure which can be useful for ``investment'' scenarios where short-term sacrifices are required to unlock long-term payoffs.

The Upper and Lower bounds account for future play by shifting baseline expectations to adjust evaluation strictness: $\hat{V}^{\text{upper}}$ strictly penalizes negative exposure, whereas $\hat{V}^{\text{lower}}$ is more tolerant of it and focuses instead on pursuing positive goals.

\section{Dynamic Programming Solver}

\begin{algorithm}[H]
\caption{\textsc{ExactSolver}}
\label{alg:exact_solver}
\begin{algorithmic}[1]
    \Statex \textbf{Input:} State $s$
    \Statex \textbf{Output:} Value vector $V(s)$ 
    \Statex \hrulefill
    
    \If{$s$ is terminal}
        \State \Return $s.\text{\textsc{GetPayoffVector}}()$
    \EndIf
    
    \State $p \gets s.\text{\textsc{Proposer}}()$
    
    \State $s_{\text{rej}} \gets \text{\textsc{RejectDeal}}(s)$
    \State $V_{\text{rej}} \gets \text{\textsc{ExactSolver}}(s_{\text{rej}})$
    
    \State $V^* \gets V_{\text{rej}}$ 
    
    \State $\mathcal{L} \gets s.\text{\textsc{GetLegalPartners}}()$
    \ForAll{$r \in \mathcal{L}$}
        \State $V_{\text{best\_with\_r}} \gets V_{\text{rej}}$
        \State $\mathcal{A} \gets \text{\textsc{GetAllJointActions}}(s, p, r)$
        
        \ForAll{$\mathbf{a} \in \mathcal{A}$}
            \State $s' \gets \text{\textsc{PlayDeal}}(s, \mathbf{a}, r)$
            \State $V_{\text{new}} \gets \text{\textsc{ExactSolver}}(s')$
            
            \If{$V_{\text{new}}[r] \geq V_{\text{rej}}[r]$}
                \If{$V_{\text{new}}[p] > V_{\text{best\_with\_r}}[p]$}
                    \State $V_{\text{best\_with\_r}} \gets V_{\text{new}}$
                \EndIf
            \EndIf
        \EndFor

        \If{$V_{\text{best\_with\_r}}[p] > V^*[p]$}
            \State $V^* \gets V_{\text{best\_with\_r}}$
        \EndIf
    \EndFor

    \State \Return $V^*$ 

\end{algorithmic}
\end{algorithm}

\section{Game Regimes and Benchmark Sweeps}
\label{app:benchmark_sweeps}

To demonstrate the expressiveness of our generator, we evaluate the three value function approximations across a broad sweep of game regimes. All games have $N=10$ players, $G=15$ goals, and an agreement budget of four actions per turn (two per player).

\textbf{Experimental Setup.~}
We consider three classes of games:
\begin{itemize}
    \item \textbf{Small generator games:} Each player has 1–2 available commitments. The approximations can be compared to the exact optimal self-interested solution (computed via Alg.~\ref{alg:exact_solver}). 
    \item \textbf{Large generator games:} Each player has 5 commitments; exact planning becomes intractable. 
    \item \textbf{Real-Life Negotiation games:} Document-grounded instances derived from real negotiation artifacts (22 players, 45 goals, $\sim$200 rounds of negotiation), providing a large and realistic testbed. 
\end{itemize}

Within each class, we isolate how key structural properties affect negotiation performance by generating 50 random instances for \emph{every} combination of five parameters:
(i) \emph{Incentive alignment}. Games are either more cooperative or more adversarial (both mixed-motive). 
(ii) \emph{Non-linearity}. The fraction of all-or-nothing goals in a game is one of $\{0.0, 0.15, 0.30, 0.50\}$. 
(iii) \emph{Goal complexity}. We use two values for the Zipf parameter $\alpha \in {1.6, 3.0}$ controlling the number of actions needed per goal. 
(iv) \emph{Latent factors}. We use two values for the rank of the player-goal matrix $d \in {5, 15}$, which indirectly sets how correlated player utilities are across issues.
(v) \emph{Payoff distribution}. Games are either balanced, positive-dominated (utilities in $[-8, 30]$), negative-dominated (utilities in $[-30, 8]$), or negative-dominated with an injected poison pill.

Evaluating the three value function approximations across these game generator sweeps was done on a 32 GB RAM machine with an Intel(R) i9-14900HX processor. Execution time took about 2 days.

\section{Training Details for Learning-Based Methods}
\label{app:learning_details}

We trained both reinforcement-learning baselines in the same staged negotiation environment used throughout the paper. Each proposer turn was decomposed into two sequential decisions: (i) partner selection and (ii) offer selection from a masked candidate set that included an explicit no-deal action. The action space was therefore stage-dependent: at the partner-selection stage, the agent chose among legal partners; at the offer-selection stage, it chose among legal candidate offers generated from the current proposer-partner pair. 

Observations combined structured per-player goal features with global game-state tensors and stage-specific masks. For each player, the feature representation included normalized goal preferences, current goal progress, remaining influence over each goal, binary- goal indicators, current goal satisfaction, whether the player could complete a goal alone, the interaction terms preference times progress, preference times satisfaction, and preference times binary-gap, as well as turn proximity and normalized remaining rounds. In addition, the model observed the current commitment matrix, the current payoff vector, the active-player mask, the acting-player identity, the selected partner identity, the turn index, turns remaining, the legal-partner mask, the legal-offer mask, and an encoded representation of the candidate offers. The environment emitted terminal-only rewards, so intermediate steps received zero reward and payoffs were realized only at the end of the negotiation. 

PPO \cite{schulman2017proximalpolicyoptimizationalgorithms} used a stage-aware actor-critic architecture. A shared context encoder first mapped the staged observation into a latent representation. This encoder was a two-layer multilayer perceptron of the form Linear-ReLU-Linear-ReLU, with hidden width 256 in both layers. The shared encoder fed three separate two-layer heads: a partner-selection head, an offer-selection head, and a scalar value head, each of the form Linear-ReLU-Linear. All hidden layers had width 256. During optimization, transitions were grouped by stage so that each minibatch used the appropriate policy head. Training used Adam with learning rate 3e-4, PPO clip parameter 0.2, value-loss coefficient 0.5, entropy coefficient 0.01, gradient clipping at norm 0.5, 4 PPO epochs per iteration, and minibatch size 64. For the fixed 10-game experiment, PPO was trained for 25 iterations with 10 episodes per iteration, corresponding to one rollout per game in each iteration. The underlying game bundle consisted of balanced adversarial 5-player games with 4 actions per player, 8 goals, and 10 proposer rounds per player. Because rewards were terminal-only and the acting player changed across decision steps, we did not use standard dense-reward return targets. Instead, each transition was labeled with the normalized terminal payoff of the player who acted at that step, and the advantage target was computed as normalized acting-player terminal payoff minus the critic prediction. 

RFE used the same staged observation and action parameterization, but replaced the actor-critic with an ensemble of stage-aware Q-networks (following the approach by \cite{zhang2024uncertaintyawarerewardfreeexplorationgeneral}). Each Q-network used the same architectural template as PPO's shared policy backbone: a two-layer shared context encoder of the form Linear-ReLU-Linear-ReLU with hidden width 256, followed by two separate two-layer Q-heads, one for partner selection and one for offer selection, each of the form Linear-ReLU-Linear. All hidden layers had width 256. Exploration was reward-free: actions were selected using an epsilon-greedy optimistic rule based on the mean predicted Q-value plus an uncertainty bonus, where uncertainty was estimated from disagreement across target-network predictions in the ensemble. The exploration ensemble was trained online from replay using Adam with learning rate 1e-4, discount factor gamma = 0.99, uncertainty coefficient beta = 1.0, epsilon = 0.2, target-network update rate tau = 0.01, variance floor 1e-4, maximum inverse-variance weight 100.0, replay capacity 50,000, warmup threshold 128 transitions, one update per environment step after warmup, and batch size 128. For the fixed 10-game experiment, we used an ensemble size of 5 and collected 100 exploration episodes on the same game bundle as PPO. 

After the reward-free exploration phase, replay transitions were relabeled with the normalized terminal payoff of the acting player, yielding an offline planning target for each decision step. A separate stage-aware planner Q-network, with the same two- layer encoder and two-layer stage heads described above, was then trained on this relabeled replay buffer using supervised regression to terminal acting-player payoff targets. This offline planner was trained with Adam using learning rate 3e-4 for 25 epochs and minibatch size 128. Final RFE evaluation used the greedy policy induced by this offline planner rather than the exploratory ensemble policy. 

Both PPO and RFE used the same staged environment, the same legal-action masking scheme, and the same fixed 10-game balanced adversarial training bundle for the comparison experiments reported in the paper. This ensured that differences between the methods reflected differences in the learning procedure rather than differences in game instances or action parameterization.

Training was done on a NVIDIA GeForce RTX 4070 (8 GB VRAM), requiring a total execution time of about 1-2 hours for both methods.

\section{Utiltiarian $\&$ Nash Social Welfare Solvers}
\label{sec:one_shot_solvers}

\paragraph{One-Shot Social Welfare Solutions.}
Instead of looking at turn-by-turn selfish negotiations, one may also want to know the absolute best possible outcome for the entire group. This ignores turn order and assumes a central planner assigns the final agreements. We look at two common objective (where $y_{i}$ is used to denote the terminal payoff of player $p_i$):
\begin{itemize}
    \item \textbf{Utilitarian Social Welfare:} Maximizes the total sum of all players' rewards ($\sum y_{i}$).
    \item \textbf{Nash Social Welfare:} Maximizes the product of all players' rewards ($\prod y_{i}$, or equivalently $\sum \log y_{i}$), which encourages fairness and prevents one player from getting nothing.
\end{itemize}

We can find these global optimums by solving a Mixed-Integer Linear Program (MILP). We optimize over the decision variables $P_{i,a} \in \{0,1\}$ (actions), $S_j \in [0,1]$ (goal satisfaction), and $y_{i} \ge \delta$ (player payoffs, bounded slightly above zero for the Nash log calculation). The formulation is:

\begin{align*}
\textbf{Maximize:} \quad 
& \sum_{i=1}^{N} y_i \quad \textbf{(Utilitarian)} \quad \text{or} \quad \sum_{i=1}^{N} \log(y_i) \quad \textbf{(Nash)} \\[5pt]
\textbf{Subject to:} \quad 
& y_i = \sum_{j=1}^{G} \mathbf{G}_{j,i} S_j \hspace{4.3cm} \text{(Player payoff sum)} \\[5pt]
& S_j = \frac{1}{|\mathcal{C}_j|} \sum_{(i,a) \in \mathcal{C}_j} P_{i,a} \hspace{2.9cm} \text{(If } j \text{ is a linear goal)} \\[5pt]
& S_j = \begin{cases} 
1 & \text{if } \sum_{(i,a) \in \mathcal{C}_j} P_{i,a} = |\mathcal{C}_j| \\ 
0 & \text{otherwise} 
\end{cases} \hspace{1cm} \text{(If } j \text{ is a all-or-nothing goal)}
\end{align*}

\section{Results}
\label{app:results}

\subsection{Reporting Uncertainty}

In our experiments, we report uncertainty using \emph{paired} standard errors across evaluation games. Since all methods are evaluated on the same set of games (shared seeds), we compute, for each game $j$, the difference in sum of payoffs relative to the No Negotiation baseline (Note: for small games, we compute the aggregate L1 error relative to the payoff of the optimal solution):
\[
d_j = x_{\mathrm{method}, j} - x_{\mathrm{no\_negotiation}, j}.
\]
This yields a set of per-game differences $\{d_j\}_{j=1}^n$ for each method, where $n$ is the number of games.

For each method, we report the mean difference
\[
\bar{d} = \frac{1}{n} \sum_{j=1}^n d_j,
\]
and the standard error of the mean
\[
\mathrm{SE}(\bar{d}) = \frac{\mathrm{std}(d_1, \dots, d_n)}{\sqrt{n}},
\]
where $\mathrm{std}(\cdot)$ denotes the sample standard deviation.

\subsection{Large games generator full results}

In Section \ref{sec:methods}, we report results on large generated games, averaged over the fraction of binary goals for compactness. We list the results for every fraction of binary goals here.

\begin{table}[htbp]
    \centering
    \resizebox{\textwidth}{!}{
    \renewcommand{\arraystretch}{1.1}
    \begin{tabular}{@{}llcccc@{}}
        \toprule
        & & \multicolumn{4}{c}{\textbf{Game Regime}} \\
        \cmidrule(l){3-6}
        \textbf{Method} & \textbf{Binary Frac.} & \textbf{Balanced} & \textbf{Negative-Dom. w/ PP} & \textbf{Negative-Dom.} & \textbf{Positive-Dom.} \\
        \midrule

        \multicolumn{6}{@{}l}{\textit{Adversarial Structure}} \\
        \midrule
        \multirow{4}{*}{Lower} 
        & 0.00 & $-26.88 \pm 9.08$ & $-189.04 \pm 11.82$ & $-182.31 \pm 13.56$ & $\mathbf{53.83 \pm 5.02}$ \\
        & 0.15 & $-33.99 \pm 9.09$ & $-193.82 \pm 11.80$ & $-183.76 \pm 13.40$ & $\mathbf{52.87 \pm 4.68}$ \\
        & 0.30 & $-36.48 \pm 9.35$ & $-185.90 \pm 11.40$ & $-176.75 \pm 12.85$ & $\mathbf{53.40 \pm 4.61}$ \\
        & 0.50 & $-39.00 \pm 9.50$ & $-217.47 \pm 12.57$ & $-200.45 \pm 13.49$ & $\mathbf{54.78 \pm 4.27}$ \\
        \addlinespace
        
        \multirow{4}{*}{Reward} 
        & 0.00 & $\mathbf{14.96 \pm 4.19}$ & $-12.56 \pm 2.82$ & $6.39 \pm 2.02$ & $47.72 \pm 4.94$ \\
        & 0.15 & $\mathbf{15.28 \pm 4.48}$ & $-11.55 \pm 3.58$ & $7.28 \pm 3.17$ & $47.34 \pm 4.57$ \\
        & 0.30 & $\mathbf{12.58 \pm 4.01}$ & $-10.54 \pm 3.24$ & $7.91 \pm 3.41$ & $45.42 \pm 4.26$ \\
        & 0.50 & $\mathbf{14.98 \pm 3.87}$ & $-16.58 \pm 3.59$ & $1.19 \pm 3.05$ & $46.94 \pm 4.22$ \\
        \addlinespace

        \multirow{4}{*}{Upper} 
        & 0.00 & $1.13 \pm 3.32$ & $\mathbf{13.90 \pm 3.00}$ & $\mathbf{16.15 \pm 2.72}$ & $-4.61 \pm 4.33$ \\
        & 0.15 & $-10.93 \pm 4.18$ & $\mathbf{14.28 \pm 3.70}$ & $\mathbf{15.86 \pm 3.83}$ & $-158.20 \pm 15.73$ \\
        & 0.30 & $-17.99 \pm 5.14$ & $\mathbf{12.16 \pm 3.12}$ & $\mathbf{11.27 \pm 3.60}$ & $-312.01 \pm 22.79$ \\
        & 0.50 & $-30.14 \pm 6.09$ & $\mathbf{5.86 \pm 3.24}$ & $\mathbf{6.29 \pm 3.65}$ & $-567.25 \pm 27.25$ \\
        \midrule
        
        \multicolumn{6}{@{}l}{\textit{Cooperative Structure}} \\
        \midrule
        \multirow{4}{*}{Lower} 
        & 0.00 & $-19.37 \pm 3.04$ & $-27.25 \pm 2.50$ & $-6.03 \pm 1.82$ & $-5.57 \pm 1.46$ \\
        & 0.15 & $-14.19 \pm 3.65$ & $-26.59 \pm 3.02$ & $-3.20 \pm 1.65$ & $-4.51 \pm 1.47$ \\
        & 0.30 & $-11.58 \pm 3.95$ & $-20.32 \pm 2.35$ & $-0.64 \pm 1.99$ & $-3.76 \pm 1.43$ \\
        & 0.50 & $-8.91 \pm 4.33$ & $-24.72 \pm 2.71$ & $-1.87 \pm 2.21$ & $-1.89 \pm 1.46$ \\
        \addlinespace
        
        \multirow{4}{*}{Reward} 
        & 0.00 & $\mathbf{4.23 \pm 1.34}$ & $-6.55 \pm 0.86$ & $\mathbf{1.85 \pm 0.78}$ & $\mathbf{1.07 \pm 0.86}$ \\
        & 0.15 & $\mathbf{6.91 \pm 2.52}$ & $-5.19 \pm 1.09$ & $\mathbf{3.45 \pm 0.98}$ & $\mathbf{2.70 \pm 0.96}$ \\
        & 0.30 & $\mathbf{11.62 \pm 3.04}$ & $-2.54 \pm 1.30$ & $\mathbf{5.42 \pm 1.21}$ & $\mathbf{3.93 \pm 1.06}$ \\
        & 0.50 & $\mathbf{13.44 \pm 3.72}$ & $-1.55 \pm 1.34$ & $\mathbf{7.01 \pm 1.43}$ & $\mathbf{4.49 \pm 1.44}$ \\
        \addlinespace

        \multirow{4}{*}{Upper} 
        & 0.00 & $-19.43 \pm 3.13$ & $\mathbf{1.26 \pm 0.79}$ & $1.25 \pm 0.85$ & $-24.66 \pm 5.01$ \\
        & 0.15 & $-34.82 \pm 5.15$ & $\mathbf{2.58 \pm 0.96}$ & $2.18 \pm 1.04$ & $-154.20 \pm 15.44$ \\
        & 0.30 & $-55.46 \pm 7.25$ & $\mathbf{3.68 \pm 1.17}$ & $4.46 \pm 1.19$ & $-303.84 \pm 22.34$ \\
        & 0.50 & $-88.40 \pm 9.24$ & $\mathbf{2.34 \pm 1.18}$ & $4.46 \pm 1.36$ & $-527.84 \pm 26.60$ \\
        
        \bottomrule
    \end{tabular}
    }
    \caption{Comprehensive breakdown of total pairwise differences ($\sum \Delta_n$) across all game regimes, incentive alignments, and fractions of all-or-nothing (all-or-nothing) goals. Values represent the Mean $\pm$ Standard Error of the Mean (SEM).}
    \label{tab:full_results_large}
\end{table}







\newpage

\newpage
\section*{NeurIPS Paper Checklist}

\begin{enumerate}

\item {\bf Claims}
    \item[] Question: Do the main claims made in the abstract and introduction accurately reflect the paper's contributions and scope?
    \item[] Answer: \answerYes{} 
    \item[] Justification: The abstract and introduction accurately describe the paper's contributions -- including the proposed benchmark, evaluation framework, and empirical analysis of negotiation methods. These claims directly match our methodological and experimental results.

\item {\bf Limitations}
    \item[] Question: Does the paper discuss the limitations of the work performed by the authors?
    \item[] Answer: \answerYes{} 
    \item[] Justification: The paper discusses several limitations, including the dependence of results on the chosen negotiation protocol and evaluation assumptions, as well as the use of perfect information and simplified interaction dynamics. We also highlight that conclusions are regime-dependent and may not generalize across different settings. These limitations are discussed throughout the paper and appendices.

\item {\bf Theory assumptions and proofs}
    \item[] Question: For each theoretical result, does the paper provide the full set of assumptions and a complete (and correct) proof?
    \item[] Answer: \answerNA{}{} 
    \item[] Justification: This paper does not include any theoretical results.

    \item {\bf Experimental result reproducibility}
    \item[] Question: Does the paper fully disclose all the information needed to reproduce the main experimental results of the paper to the extent that it affects the main claims and/or conclusions of the paper (regardless of whether the code and data are provided or not)?
    \item[] Answer: \answerYes{} 
    \item[] Justification: We provide detailed information necessary to reproduce our main results. Specifically, we describe the configuration of our game generator sweeps in Appendix~\ref{app:benchmark_sweeps}, and we provide full details of the training procedures, hyperparameters, and implementation choices for the PPO and RFE baselines in Appendix~\ref{app:learning_details}. We further provide our code at an anonymized repository: \url{https://anonymous.4open.science/r/negotiation_MARL-46B8/}.

\item {\bf Open access to data and code}
    \item[] Question: Does the paper provide open access to the data and code, with sufficient instructions to faithfully reproduce the main experimental results, as described in supplemental material?
    \item[] Answer: \answerYes{} 
    \item[] Justification:  We provide our code at an anonymized repository: \url{https://anonymous.4open.science/r/negotiation_MARL-46B8/}. We also provide detailed information necessary to reproduce our main results (see the answer to question 4 in the checklist).
    \item[] Guidelines:

\item {\bf Experimental setting/details}
    \item[] Question: Does the paper specify all the training and test details (e.g., data splits, hyperparameters, how they were chosen, type of optimizer) necessary to understand the results?
    \item[] Answer: \answerYes{} 
    \item[] Justification: We provide details about our training/testing setup for the RL-baseline methods in Appendix~\ref{app:learning_details}.

\item {\bf Experiment statistical significance}
    \item[] Question: Does the paper report error bars suitably and correctly defined or other appropriate information about the statistical significance of the experiments?
    \item[] Answer: \answerYes{} 
    \item[] Justification: We report uncertainty using standard errors of the mean, computed over evaluation games. Specifically, we use a paired evaluation setup (shared seeds across methods) and compute per-game differences relative to a baseline, capturing variability across games. Uncertainty measures correspond to the standard error of the mean of these differences. We provide full details of this procedure, including formulas and assumptions, in Appendix~\ref{app:results}.

\item {\bf Experiments compute resources}
    \item[] Question: For each experiment, does the paper provide sufficient information on the computer resources (type of compute workers, memory, time of execution) needed to reproduce the experiments?
    \item[] Answer: \answerYes{} 
    \item[] Justification: We provide detailed information about compute resources and execution time for each experiment in the relevant sections of the paper. Specifically, we report the hardware used (including GPU/CPU type and memory) and the corresponding execution times for both training runs (Appendix~\ref{app:learning_details} and large-scale evaluation sweeps (Appendix~\ref{app:benchmark_sweeps}).
    
\item {\bf Code of ethics}
    \item[] Question: Does the research conducted in the paper conform, in every respect, with the NeurIPS Code of Ethics \url{https://neurips.cc/public/EthicsGuidelines}?
    \item[] Answer: \answerYes{} 
    \item[] Justification: This work complies with the NeurIPS Code of Ethics. It does not involve human subjects or sensitive personal data; document inputs are derived from role-play materials and used in non-identifying form. The work poses minimal safety or misuse risks, and we provide sufficient detail to ensure transparency and reproducibility.

\item {\bf Broader impacts}
    \item[] Question: Does the paper discuss both potential positive societal impacts and negative societal impacts of the work performed?
    \item[] Answer: \answerYes{} 
    \item[] Justification: We discuss both positive and negative societal impacts in the conclusion, including potential applications for supporting high-stakes decision-making and risks of misuse in adversarial contexts.
    
\item {\bf Safeguards}
    \item[] Question: Does the paper describe safeguards that have been put in place for responsible release of data or models that have a high risk for misuse (e.g., pre-trained language models, image generators, or scraped datasets)?
    \item[] Answer: \answerNA{} 
    \item[] Justification: Our paper poses no such risks.

\item {\bf Licenses for existing assets}
    \item[] Question: Are the creators or original owners of assets (e.g., code, data, models), used in the paper, properly credited and are the license and terms of use explicitly mentioned and properly respected?
    \item[] Answer: \answerYes{} 
    \item[] Justification: The position papers used in this work were obtained from a structured negotiation exercise for academic use. To respect usage restrictions, we do not redistribute the original documents and instead release only processed, non-identifying game representations. The original materials are included in the supplementary submission for review purposes only.

\item {\bf New assets}
    \item[] Question: Are new assets introduced in the paper well documented and is the documentation provided alongside the assets?
    \item[] Answer: \answerYes{} 
    \item[] Justification: We introduce a new benchmark and associated game instances derived from negotiation documents. We provide code for game generation (in anonymized form) and include detailed documentation of the data schema, generation process, and evaluation setup in the paper and appendices, enabling understanding and reuse of the assets.

\item {\bf Crowdsourcing and research with human subjects}
    \item[] Question: For crowdsourcing experiments and research with human subjects, does the paper include the full text of instructions given to participants and screenshots, if applicable, as well as details about compensation (if any)? 
    \item[] Answer: \answerNA{} 
    \item[] Justification: The paper does not involve crowdsourcing nor research with human subjects.

\item {\bf Institutional review board (IRB) approvals or equivalent for research with human subjects}
    \item[] Question: Does the paper describe potential risks incurred by study participants, whether such risks were disclosed to the subjects, and whether Institutional Review Board (IRB) approvals (or an equivalent approval/review based on the requirements of your country or institution) were obtained?
    \item[] Answer: \answerNA{} 
    \item[] Justification: The paper does not involve crowdsourcing nor research with human subjects.

\item {\bf Declaration of LLM usage}
    \item[] Question: Does the paper describe the usage of LLMs if it is an important, original, or non-standard component of the core methods in this research? Note that if the LLM is used only for writing, editing, or formatting purposes and does \emph{not} impact the core methodology, scientific rigor, or originality of the research, declaration is not required.
    \item[] Answer: \answerYes{} 
    \item[] Justification: We detail our LLM-based methodology in Appendix~\ref{app:topfile_creation}.

\end{enumerate}

\end{document}